# Unsupervised and semi-supervised clustering methods to identify and refine participant experience levels in educational research

Julien-Pooya Weihs[1,2,†] *(julien-pooya.weihs@uib.no)*, Adrien Weihs[3] *(weihs@math.ucla.edu)*,
Vegard Gjerde[4] *(vegard.gjerde@uib.no)*, Helge Drange[1,2] *(helge.drange@uib.no)*

[1] Geophysical Institute, University of Bergen, Norway
[2] Bjerknes Centre for Climate Research, Bergen, Norway
[3] Department of Mathematics, University of California Los Angeles, USA
[4] Department of Physics and Technology, University of Bergen, Norway
[†] corresponding author

## Abstract

The progression from novice to disciplinary expert is a longstanding area of inquiry in educational research. Studies investigating such progressions have often resorted to participants' self-assessments or other qualitative indicators as a starting point to define experience. But does a participant's estimated experience coincide with metrics derived from their conceptual understanding of a discipline? Using data extracted from over 150 concept maps, we first demonstrate that disciplinary experience is a reliable variable to explain differences in conceptual understanding across a highly diverse learners' population. Through a comparison of unsupervised and semi-supervised models, we then motivate clustering participants into three distinguished experience levels, and support such a classification performed in other studies of educational research. By analysing cluster composition, we also identify discrepancies between the perceived and predicted experience levels of the study participants. Lastly, for studies processing participants data through network analysis, we present insights into statistically significant metrics that can characterise each experience level, and advocate for the use of node-level metrics in such studies.

## Introduction

With the widespread digitalisation of societal processes, the number of large and complex datasets describing both human and non-human systems is rapidly increasing. A central methodological challenge in analysing such data is to identify underlying structural patterns that organise it at a higher level (Williamson, 2017). Uncovering such patterns is essential for reducing the data's complexity, facilitating its interpretation, and informing both theoretical insight and practical applications.

In educational research, one increasingly common method for uncovering higher-order structure in data is clustering, a form of unsupervised learning that partitions it into internally similar, externally dissimilar groups (Vellido et al., 2010). Clustering techniques are particularly well suited for exploring patterns such as learner behaviour (Delgado et al., 2021; Peach et al., 2019; Prevett et al., 2021), student profiles (Darcan & Badur, 2012; Gorard & Cheng, 2011; Liu et al., 2022), or knowledge structures (He et al., 2023; Mineau & Godin, 1995).

Recent advances have introduced graph-based representations of learner knowledge structures, where nodes represent concepts and edges represent perceived relationships (Koponen & Mäntylä, 2020; Thurn et al., 2020; Wagner & Priemer, 2023). As an example, concept maps describing scientific phenomena, when formalised as graphs, can therefore be analysed using tools from network science. This allows researchers to compute graph- and node-level metrics that reflect the structural complexity and organisation of individual learners' conceptual understanding.

While clustering methods have been applied to such graph-derived data, most studies remain limited in scope. For instance, He et al. (2023) applied clustering to graphs using structural indicators (e.g., number of nodes and branches), without leveraging the relational structure captured in node-level metrics. Moreover, most studies have not rigorously connected results to learners' background characteristics, such as disciplinary experience. Instead, they tend to focus on student grades and performance, as described by Albreiki et al. (2024) and Ebli (2024). Lastly, although semi-supervised classification approaches offer a promising middle ground between knowledge-driven grouping and data-driven clustering, they remain underexplored in educational research (Karlos et al., 2020; Kostopoulos & Kotsiantis, 2022).

Our study addresses these gaps by combining unsupervised and semi-supervised clustering with graph-based analysis of concept maps to explore conceptual understanding in science education. These methods are not new, but combining them represents a novel approach in educational research. We illustrate our approach using data from atmospheric sciences, though the methodology is broadly applicable across STEM domains. To do so, we investigate whether learners' self-assessed disciplinary experience reliably predicts their actual conceptual understanding, as reflected in their concept maps. We situate our perspective on conceptual understanding within the framework developed by Kvanvig (2003) and Elgin (2012), in which understanding relates to both items of information (such as scientific concepts) and the ways to interconnect them (such as causal relationships between concepts).

We begin by using an analysis of variance to assess which learner characteristics best explains differences in conceptual understanding. We then examine how learner graphs group together based on their structural and internal properties, applying unsupervised clustering to identify similarities in graph network metrics. Following this, we evaluate how partial expert-informed labelling – such as assigning experience levels to a subset of concept maps – can enhance classification of the full dataset using semi-supervised learning (SSL). Finally, we analyse which graph-derived features most distinctly characterize different levels of experience, offering interpretable insights into the development of conceptual understanding in a discipline.

Our study is organized around the following research questions:
1. Among learners' self-assessed experience, academic level, and disciplinary background, which of these characteristics best explains variance in their graphs?
2. Which learner graphs exhibit similarities in terms of structural graph- and node-level properties? Do learners of similar experience level produce graphs which exhibit similarities in terms of structural graph- and node-level properties?
3. To what extent can partial information about learners' experience improve classification into experience levels?
4. Which graph features explain the most variance between experience-based clusters?

## Methods

### Setting and Population

This study draws on the dataset published by Weihs, Gjerde, et al. (2025), consisting of 153 hand-drawn concept maps. Each map included unlabelled directed edges and nodes representing scientific concepts and processes. The data were collected between Nov 2022 and July 2024, during structured concept-mapping exercises. Participants were asked to graphically describe the life-cycle of a cloud, using pen and paper. They were given approximately 10-15 minutes to complete the task. The concept maps were thematically analysed (Clarke & Braun, 2017) to identify key concepts and relationships, and subsequently converted into directed graph networks. Example graph networks converted from the collected concept maps, along with their computed metrics (described in the next subsection), are illustrated in Figure 1.

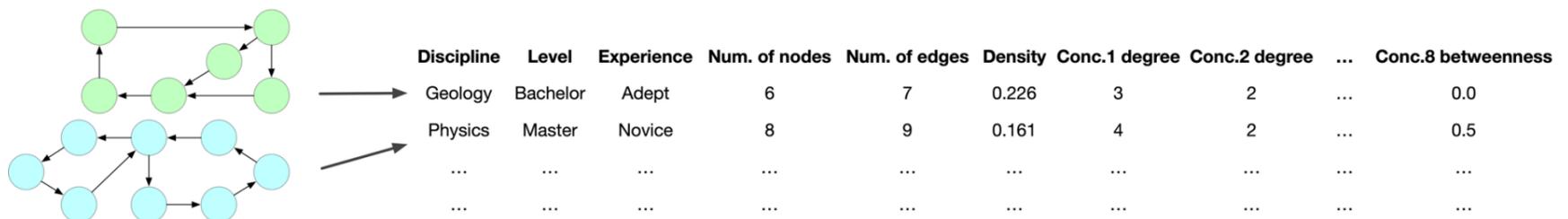

*Figure 1: Example transformation of graphs (derived from hand-drawn participant concept maps) to a general table of metrics, based on the analysis presented in Weihs, Gjerde, et al. (2025). 'Conc.x' designates an arbitrary 'Concept x'.*

The participants were affiliated with five universities across Norway and Switzerland, covering a broad spectrum of STEM disciplines. Their academic level ranged from early bachelor students to senior researchers with decades of research activity in their respective fields. Importantly, the participants' academic exposure to cloud physics varied widely: some had *'barely ever heard'* of the discipline, while others were internationally recognised experts in the field.

Self-reported experience with cloud physics was collected from the participants in the form of short free-text responses, ranging from single words (e.g., *'None'*, *'Lots'*) to more elaborate statements (e.g., *'Cloud physics has been a topic in two or three courses'*, *'One bachelor course and some cloud parametrisation research'*). Following inductive thematic analysis (Clarke & Braun, 2017), Weihs, Gjerde, et al. (2025) categorised the responses into four experience levels: *Novice*, *Adept*, *Proficient*, and *Expert*. For certain analyses, the groups were also merged into broader groups: *Beginner* (*Novice* + *Adept*) and *Advanced* (*Proficient* + *Expert*).

### Research design

This subsection details all the methods used to preprocess and cluster participants' data represented by graphs, based on structural and metric similarities. The routines include metrics computation, feature selection, clustering, and cluster evaluation. We also briefly introduce the main idea behind an analysis of variance. The combination of all presented methods into a full work pipeline is presented in the next subsection and Figure 2.

*Graph metrics computation and scaling*

For each participant, a directed graph was constructed from their collected concept map, as described by Weihs, Gjerde, et al. (2025): nodes represent scientific concepts, and edges indicate relations between them, according to the participants. These graphs are characterised by mathematical properties such as their size, connectedness, or density. Furthermore, we attribute centrality scores to each node, reflecting different importance criteria they verify, as described by Newman (2001). We computed the following usual metrics used in network analysis for each graph:
- Graph-level metrics (for the entire graph)
    - Number of nodes and edges (basic size indicators)
    - Density (the ratio of existing edges to all virtually possible edges)
    - The number of shortest paths between all possible pairs of nodes
    - Average shortest path length between all possible pairs of nodes
    - Diameter (the length of the longest shortest path in the graph)
    - Intertwinement (the ratio of the diameter to the number of nodes)
    - The number of strongly connected components (subgraphs of mutually reachable nodes)
- Node-level metrics (for each concept)
    - Degree centrality (the fraction of all nodes the node is connected to)
    - In-degree (the number of edges pointing to the node)
    - Out-degree (the number of edges pointing out of the node)
    - Betweenness centrality (the extent to which a node lies on shortest paths between others)
    - Left eigenvector centrality (capturing the influence from incoming nodes)
    - Right eigenvector centrality (capturing the influences from outgoing nodes)

All these numerical metrics were scaled using min-max normalization (Pedregosa et al., 2011) to the $[0,1]$ interval, where the largest occurring value of a metric corresponds to '1' and the smallest to '0'.

*Analysis of variance*

The computation of graph metrics provides quantitative characteristics for each participant's graph, which can vary largely across the dataset. To determine whether academic level, field of activity, or disciplinary experience in cloud physics most strongly accounts for these differences, we apply an analysis of variance, ANOVA (Scheffe, 1999). ANOVA is a statistical method that compares the means of a numerical variable across two or more categorical groups, testing whether the observed differences are statistically significant beyond what would be expected by random variation. The test relies on $F$-statistics for each feature, which compute the ratio of between-group variance (how much the group means deviate from the global mean) to within-group variance (how much individual values deviate from their group mean). We conducted ANOVA across all numerical graph features, producing a set of $F$-values for each categorical grouping variable, as presented in Figure 3. The grouping variable associated with the highest mean $F$-values across features is interpreted as the one that best differentiates learners' graph characteristics.

*Feature selection*
Although the computed graph metrics describe a wide range of structural and internal properties, not all of them are equally informative for distinguishing or comparing graphs. For example, node-level metrics associated with concepts that appear in one graph will have zero values in all other graphs where the concept is absent, limiting their usefulness for comparison. Furthermore, whether by construction of the concept maps or through the effects of increased disciplinary experience, some metrics could vary together with underlying correlation; for instance, changes in betweenness centrality for a given concept may consistently co-occur with changes in its in-degree. In such cases, the metrics could be considered redundant, and retaining both would not add explanatory power.

Feature selection (Guyon & Elisseeff, 2003) was performed to retain the most significant features characterizing clusters of participants. It reduces the noise and redundancy in the data, making the clustering faster and more reliable, and improves interpretability by rendering clearer cluster patterns. In our educational context, feature selection ensures that the retained metrics represent distinct and informative dimensions of learner's conceptual structures. The feature selection process is performed in two steps:
- First, we applied a variance threshold (denoted by $\sigma^2$) to remove features with minimal variation across participants, as such features lack discriminative power and contribute little to meaningful data partitioning. For example, node-level centralities of rarely used concepts often yield low variance. In contrast, features with higher variance are more likely to capture meaningful differences between participants. Accordingly, features with variance below a predefined threshold (typically, $\sigma^2 < 0.001$) were discarded.
- Second, we performed correlation-based feature pruning to remove features that were highly collinear. Redundant features introduce multicollinearity, which can bias clustering results. We used the absolute Pearson correlation method (Pearson, 1895) to compute the correlation coefficient $|\rho_{jk}|$ (denoted $\rho$ in the rest of this study) between all pairs of retained features, and removed one feature from each pair with correlation values above a set threshold (typically, $\rho > 0.75$). Such an empirically chosen threshold aligns with standard practice in high-dimensional feature selection to control redundancy without overly compromising representativeness.

*Data clustering on reduced set of features*
Once the reduced set of significant features is defined, a clustering algorithm can be run to partition the participant graphs into clusters. However, clustering algorithms vary in their assumptions about data structure and similarity, and several common algorithms were considered. For a broader comparison of these methods and other clustering algorithms, see Rui and Wunsch (2005) and Rodriguez et al. (2019). These include *hierarchical clustering, K-means*, *Gaussian Mixture Models*, and *spectral clustering* (Pedregosa et al., 2011).
*Hierarchical clustering*, in particular *agglomerative clustering* (Murtagh & Contreras, 2012), builds a tree-like structure of the data and is useful for exploring relationships in small datasets, using a bottom-up approach that progressively merges clusters (and groups thereof) together with increasingly looser similarity criteria.

Centroid-based grouping methods, such as *K-means*, also referred to as Lloyd's algorithm (Lloyd, 1982), and *Gaussian Mixture Models* (Reynolds, 2015), operate under the assumption that a small number of central points, or *prototypes*, can adequately represent the data. Each data point is then assigned to its nearest prototype based on simple metrics such as the Euclidean distance. However, it also assumes that all points within a cluster lie near a central prototype – an assumption often violated in practice, as real-world data frequently exhibit complex, non-linear structures that are poorly captured by centrally-defined models: *K-means* is most efficient when the number of groups is known and the data points are well separated, but is sensitive to outliers, while *Gaussian Mixture Models* handle clusters of flexible shapes and varying densities, but assume gaussian distribution for the data, and are sensitive to parameter initialisation.

*Spectral clustering* (Von Luxburg, 2007) allows to capture subtle and irregular groupings that traditional distance-based clustering may overlook. Indeed, this method groups data based on pairwise similarities and allows for transitive propagation of similarity: if point A is similar to point B, and point B to point C, then A and C may be grouped together – even without being in direct proximity of each other. Furthermore, spectral clustering supports flexible similarity definitions, making it well-suited for contexts where standard distance metrics are inadequate. We therefore selected spectral clustering as the primary algorithm for grouping participants based on the structural properties of graphs, using a Gaussian kernel with the Euclidean distance to define pairwise similarity. This similarity is used as a basis to construct a k-nearest-neighbours graph (Von Luxburg, 2007) over our features. In this graph, vertices represent participants, and edges are added between each node and its *k* nearest neighbours in feature space, with edge weights reflecting local similarity. This design emphasizes local structure and limits long-range interactions, which is consistent with the homophily assumption: if two participants belong to the same experience group, the structural features of their concept maps should be similar, resulting in a larger edge weight between them in the graph and, ultimately, a shared cluster label. This aligns with our research objective in RQ2, namely investigating whether, if two concept maps are similar, they are drawn by learners of similar experience. We also emphasize that the use of graphs in our method is algorithmic and distinct from the graph-like nature of the original data. While our dataset consists of concept maps, these are first transformed into high-dimensional feature vectors; the latter are used to construct the graph used as input for the spectral clustering, as well as the Laplace and Poisson learning (see next subsection).

Although rarely applied in educational research, spectral clustering has shown promise in a few related studies. For instance, clusters of learners have been identified by investigating the relationship between either screen time (Hawi & Samaha, 2017) or behaviours in social networks (Obadi et al., 2010) and their academic performance, and other applications of spectral clustering in education have included curriculum design (Zhou et al., 2022). A known limitation of spectral clustering, and pairwise grouping methods more broadly, is the computational burden of evaluating all pairwise similarities, which scale quadratically with the number of samples (Yom-Tov & Slonim, 2009). However, our dataset consists of only 153 samples, each representing the structural features of a graph, making the computational cost negligible.
Spectral clustering requires the number of desired clusters, or the dimension of the projection subspace (Pedregosa et al., 2011), to be input. In connection with our dataset, this would mean investigating the graphs through either three (*Novice, Adept, Advanced*) or four (*Novice, Adept, Proficient, Expert*) experience levels. We test our analysis for both cases, which yields a chosen number of clusters $c = 3,4$.

*Clustering accuracy and robustness*
While clustering algorithms generate groupings based on statistical criteria, it is also important to evaluate how well these groupings correspond to known or expected structures in the data. A common approach (Calder, 2022) is to compute an accuracy score by finding the best possible alignment between predicted (cluster) labels and true (grouping) labels, using a linear sum assignment. This score reports the proportion of correctly clustered items under the best label permutation. Other common accuracy metrics include the Adjusted Rand Index, ARI (Hubert & Arabie, 1985), and the Normalized Mutual Information, NMI (Cover & Thomas, 1991). ARI evaluates how similar the clusters and groups are by checking how many pairs of items are consistently grouped or separated in both the predicted clusters and true groups. NMI measures how much knowing one predicted cluster label informs about a group label, considering distribution overlaps between clusters and groups. In this study, we will mostly focus on the accuracy score, but also compute the ARI and NMI for complementarity, as displayed in Figure 4 and Figure 5.

Another critical aspect of clustering analyses is robustness, or how sensitive the output is to a variation in input parameters, such as here the variance threshold $\sigma^2$ and the correlation threshold $\rho$. By varying $\sigma^2$ and $\rho$ across ranges, generating clusters and measuring their accuracy, we perform a first sensitivity analysis of the clustering accuracy, captured in a heatmap, see Figure 4 and Figure 5. Similarly, a clustering output can also be compared across models, to verify its behaviour in a parametrised domain of interest. We perform a second type of sensitivity analysis by selecting a favoured value of $\rho$ and comparing the accuracy scores as a function of $\sigma^2$ and for different clustering models, leading to a clustering algorithm comparison, see Figure 6.

*Semi-supervised clustering*
The motivation behind using unsupervised clustering is to test whether participants' graphs reveal more statistically coherent groupings than those based on self-reported experience alone. One underlying hypothesis of this process is that there is no *a priori* knowledge of what characterises a 'Novice-map' or 'Expert-map'. Only an *a posteriori* assessment of the clusters' most significant features could answer this point. We however presume that a disciplinary expert would be able to rate the content quality of the collected concept maps, leading us to the assumption that by involving such an expert, we could gain confirmation on a subset of graphs as belonging to either of the experience levels. This is equivalent to positing that a certain percentage, also called label rate $r$, of the grouping labels per experience levels are correct. If their labels are correct, these graphs carry information about characteristics specific to their respective experience groups. By leveraging the geometrical relationships between labelled and unlabelled concepts maps in our feature space, our objective is to propagate label information and extend group assignments to the entire dataset. This setting is also known as the SSL paradigm in machine learning.

In this study, we perform SSL clustering by comparing Laplace learning (Zhu et al., 2003) and Poisson learning (Calder et al., 2020). Both approaches are graph-based, meaning they operate on data structured as a graph, and we rely on the same graph construction as for our clustering study detailed in the previous section. While the core mechanism of these SSL methods involves propagating labels across the graph's edges, Poisson learning is particularly well-suited to sparse label regimes where $r$ is small. We also remark that our choice of graph-based learning algorithms is specifically tailored to the SSL setting. Indeed, the graph representation of our features provides a natural and principled way to relate labelled and unlabelled concept maps.

## Full pipeline

### Structure
Based on the methods described above, our work pipeline is outlined in Figure 2, with a more extensive and detailed version of it in the appendix (Figure 10). The analysis has been performed in Python, mainly using the *scikit-learn* (Pedregosa et al., 2011) for the unsupervised methods and *graphlearning* (Calder, 2022) for the SSL algorithms. We address our research questions in sequential steps: RQ1 is answered through ANOVA analysis, RQ2 is answered by the final output of the unsupervised analysis, RQ3 is answered by a comparison of the unsupervised and SSL outputs, and RQ4 is answer by investigating the cluster compositions of the SSL method.

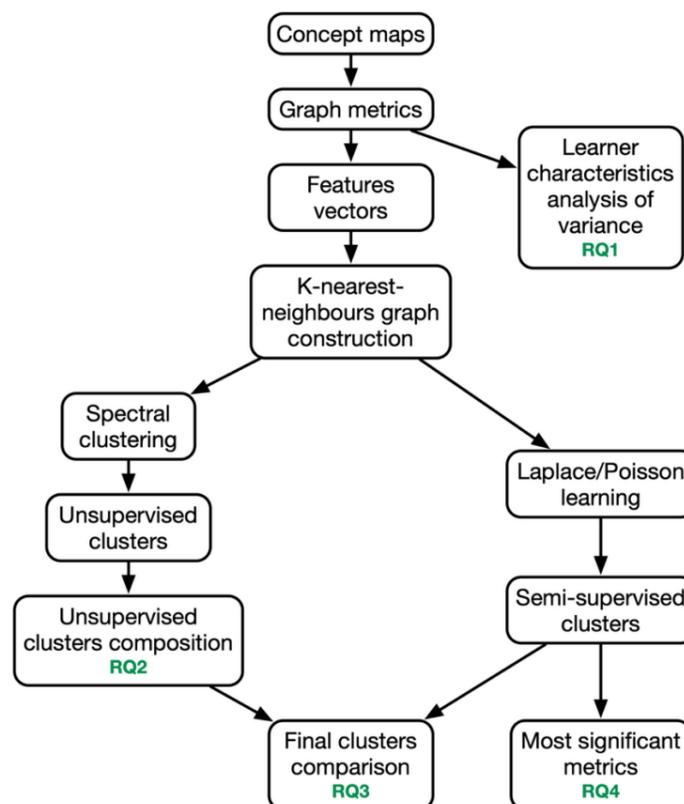

*Figure 2: Work pipeline describing the unsupervised and semi-supervised clustering analyses of the participants data according to various graph- and node-level metrics. The outputs addressing the four research questions of this study are highlighted with the mentions 'RQx' (in green).*

## Results
An interpretation of the results is provided in the further Discussion section.

### Analysis of variance
Figure 3 presents the ANOVA mean $F$-statistic for all three categorical independent variables of the dataset. The p-values of each variable test are below the commonly adopted $p < 0.05$ threshold of statistical significance (corresponding to a 95% likelihood of rejecting the null-hypothesis) for all three variables. Additionally, the 'discipline' and 'experience' variables show scores of 0.001, making their respective mean $F$-values extremely significant.

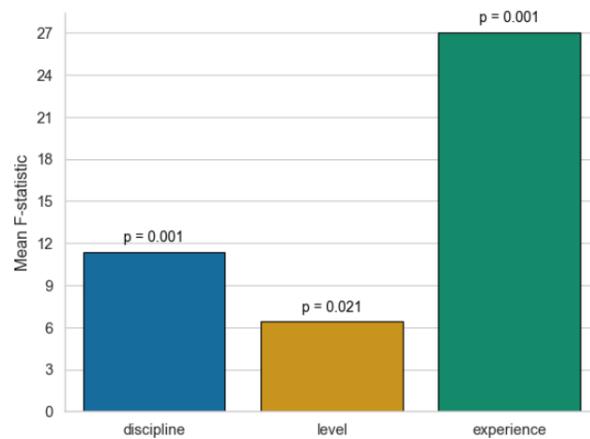

*Figure 3: ANOVA mean F-statistic by grouping variables 'discipline', 'level', and 'experience'. The p-values for each variable statistic are given over each bar.*

## Unsupervised spectral clustering

Figure 4 and Figure 5 display the first part of the sensitivity analysis, presenting accuracy, ARI and NMI scores for unsupervised clusters. Figure 4 shows results for a 4-clusters analysis based on the initial experience levels *Novice*, *Adept*, *Proficient*, and *Expert*. Figure 5 shows results for a 3-clusters analysis based on the merged levels *Novice*, *Adept*, and *Advanced*.

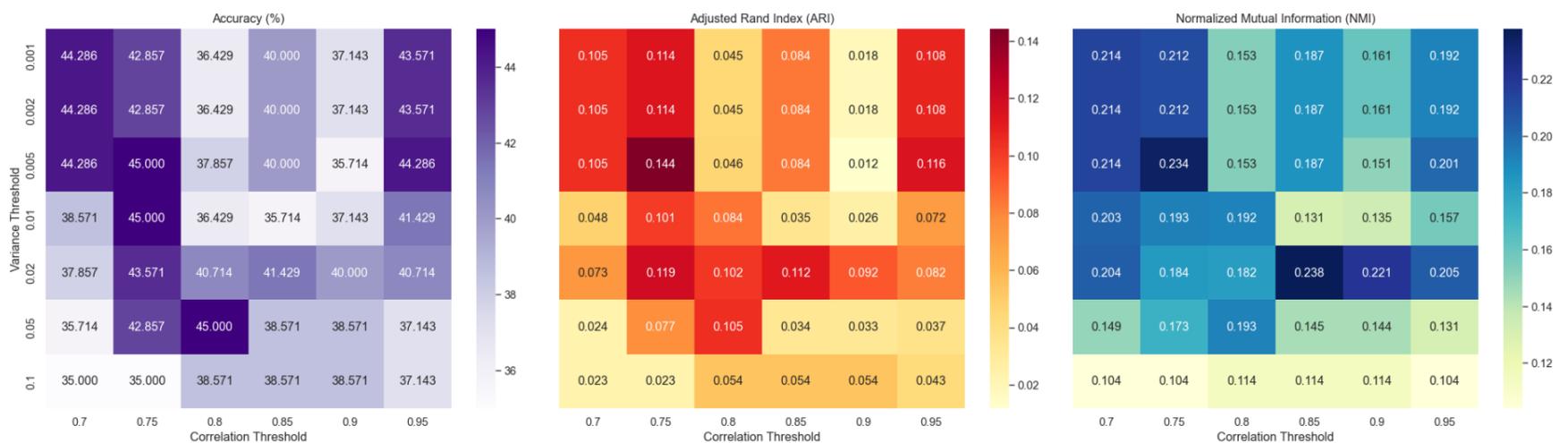

*Figure 4: Clustering accuracy (left, in purples), ARI (middle, in reds), and NMI (right, in blues) scores for 4 clusters (starting from Novice, Adept, Proficient, Expert) generated through unsupervised spectral clustering. The heatmaps all feature the same value ranges: [0.7-0.95] for the correlation threshold $\rho$ on the x-axis, and [0.001-0.1] for the variance threshold $\sigma^2$ on the y-axis.*

In the 4-cluster analysis (Figure 4), the highest accuracy values are obtained for $\rho_{4\text{-}max} = 0.75$ and $\sigma^2_{4\text{-}max} \in \{0.005, 0.01\}$, where we use the symbol '∈' to indicate that the parameter can take either of the values of the set. Among those, $\sigma^2_{4\text{-}max} = 0.005$ yields the highest ARI and NMI scores. Therefore, the optimal threshold pair for the 4-clusters case is chosen as (0.75, 0.005), retaining 22.3% (185 out of 830, not shown in the figure) of the initial features for the algorithmic computations.

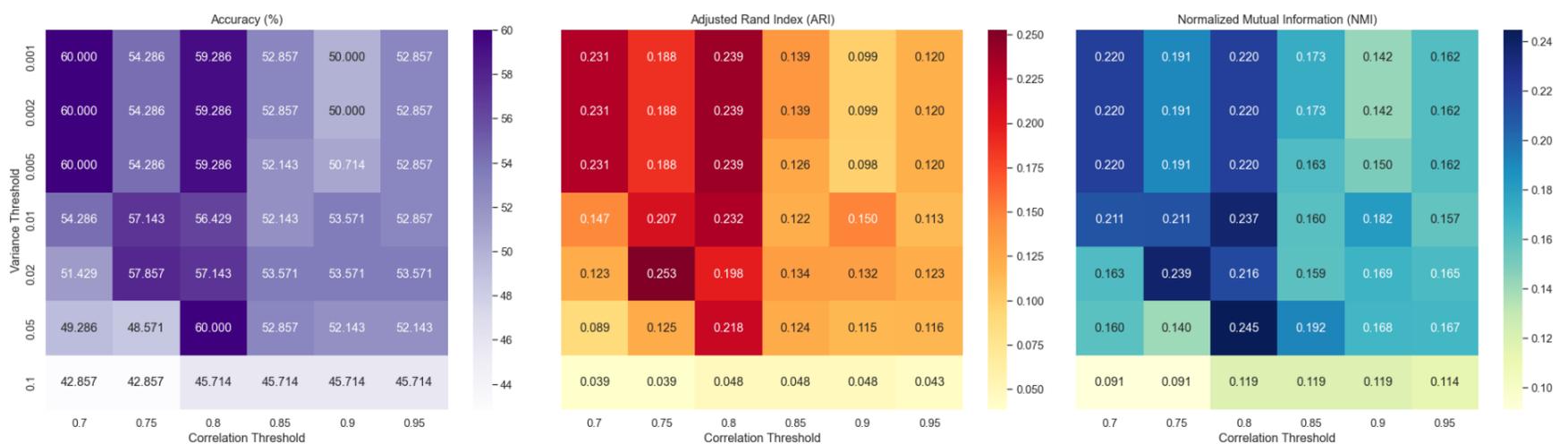

*Figure 5: Same as Figure 4, but for an analysis with 3 clusters (starting from Novice, Adept, Advanced).*

For the 3-cluster analysis (Figure 5), the highest accuracy is reached for two regions: (i) $\rho_{3\text{-}max\text{-}1} = 0.70$ with $\sigma^2_{3\text{-}max\text{-}1} \in \{0.001, 0.002, 0.005\}$, and (ii) $\rho_{3\text{-}max\text{-}2} = 0.80$ with $\sigma^2_{3\text{-}max\text{-}2} \in \{0.001, 0.002, 0.005, 0.05\}$. To maximise accuracy at 60%, we select $\rho_{3\text{-}max} = 0.70$. Based on the stability of neighbouring points, $\sigma^2_{3\text{-}max} = 0.002$ is chosen, giving the most robust plateau of high performance. The selected thresholds (0.70, 0.002) retain 17,7% (147 out of 830, not shown in the figure) of the initial features. The robustness of this pair is similar across all three metrics: equal values for neighbouring points along the y-axis, limited decrease of the metrics for $\sigma^2_3 = 0.75$, and similar values at $\sigma^2_3 = 0.80$.

Overall, the 3-clusters analysis yields better performance: the highest accuracy (60%) and ARI (0.253) exceed the respective values from the 4-clusters analysis (45% and 0.144). This motivates a continued focus on the 3-clusters analysis in the remainder of this study. Comparisons with the 4-cluster analysis is provided whenever relevant in the Discussion section.

The second part of the sensitivity analysis is shown in Figure 6. It compares clustering accuracy across four algorithms under the 3-clusters configuration with a fixed correlation threshold value of $\rho_{3\text{-}max} = 0.70$, as motivated above. Spectral clustering outperforms the other models in the low-variance threshold region [0.001-0.005] before its accuracy declines across the rest of the considered range.

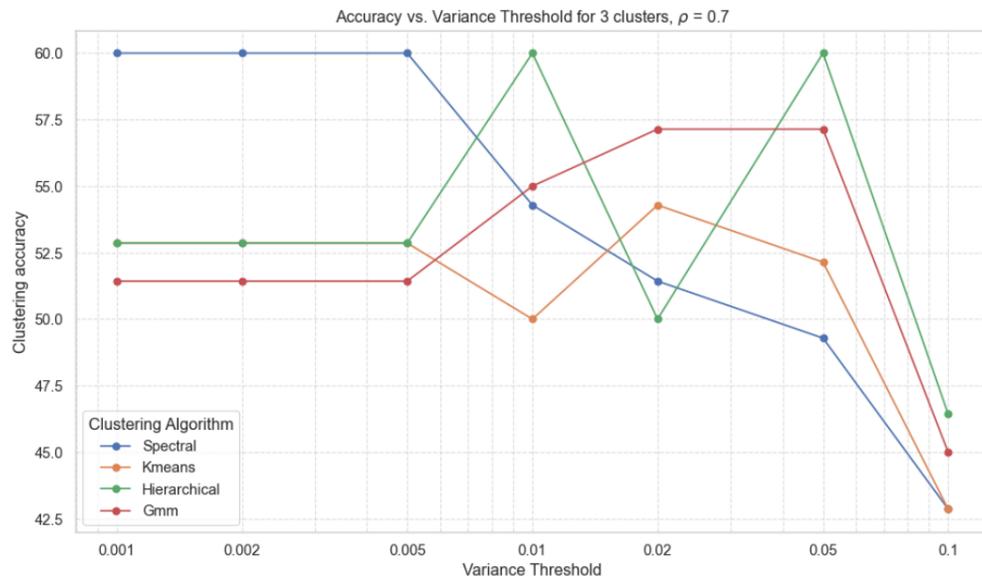

*Figure 6: Clustering accuracy of Spectral, K-means, Hierarchical and Gaussian mixture model (Gmm) algorithms computed for 3 clusters across a range of variance threshold values $\sigma^2$ [0.001-0.1], and for a fixed correlation value $\rho_{3\text{-}max} = 0.70$.*

Figure 7 presents the composition of three clusters generated by unsupervised spectral clustering under two different experience labelling schemes and threshold pairs. As an unsupervised method, the clustering is exploratory in nature: the algorithm assigns participants to clusters solely based on graph-derived features, without access to their experience labels. These labels are used only *post hoc* to interpret and evaluate the cluster composition.

The left panel shows results for the reduced three experience levels *Novice*, *Adept*, *Advanced*, clustered using the threshold $\rho_{3\text{-}max} = 0.70$ and $\sigma^2_{3\text{-}max} = 0.002$. Cluster 2 is the largest and predominantly composed of *Advanced* learners, while clusters 0 and 1 are similar in composition, each containing a comparable mix of *Novice* and *Adept* learners. Cluster 1 is about half the size of cluster 2, and cluster 0 falls in size between them. The right panel shows the clustering composition using the four initial experience levels *Novice*, *Adept*, *Advanced*, *Expert*, and the optimised threshold pair (0.80, 0.02). Here, cluster 2 consists mostly of *Expert* and *Proficient* learners, cluster 0 is dominated by *Novice* and *Adept* learners, and cluster 1 contains a more heterogeneous mix of all four experience levels. Unlike the left panel, the clusters in the right panel are more balanced in size.

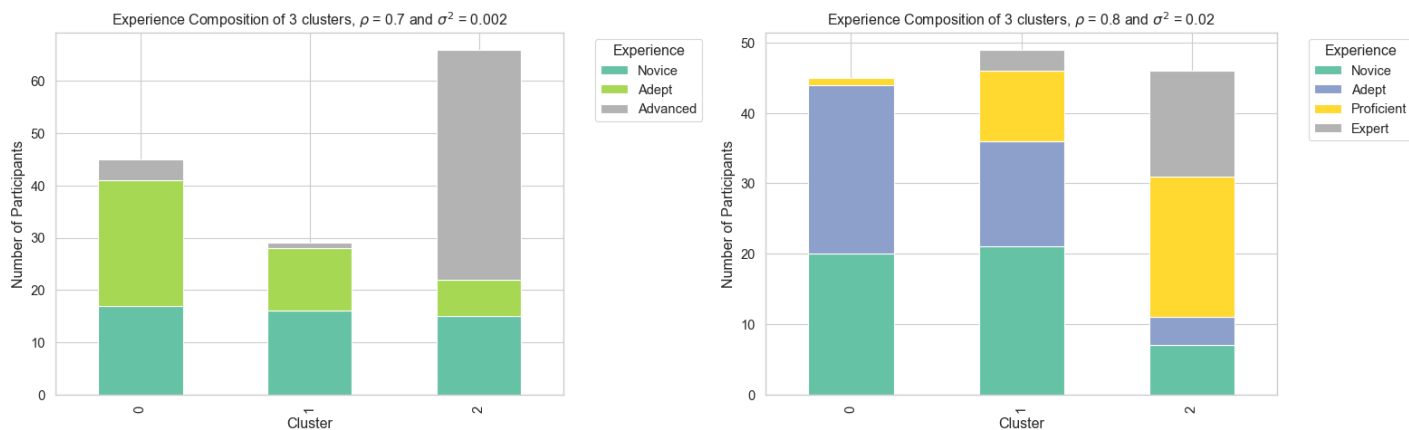

*Figure 7: Composition of 3 clusters generated by unsupervised spectral clustering under two different experience groupings. Left: using three experience levels (Novice, Adept, Advanced), with thresholds $\rho_{3\text{-}max} = 0.70$ and $\sigma^2_{3\text{-}max} = 0.002$. Right: using four experience levels (Novice, Adept, Proficient, Expert), with thresholds $\rho = 0.80$ and $\sigma^2 = 0.02$. Bar heights represent the number of participants per experience level within each predicted cluster.*

### Semi-supervised clustering

In Figure 8, we show the evolution of mean clustering accuracy scores for label propagation experiments within the SSL routines. The experiments are run across a range of varying parameters $r$ and $k$, and for both the Poisson and Laplace learning models. Until a label rate around $r = 0.5$, the Poisson model performs better than the Laplace model. Afterwards, the Laplace model surpasses the Poisson model and increases in accuracy until the label rate value of $r_{max} = 0.7$. At $r_{max}$, we tested the maximal accuracy across a wide range of $k$ values (only partially shown in the figure), and noted no statistically significant difference, making us pick $k_{max} = 38$. The Poisson model remain roughly constant at all label rates $r > 0.1$.

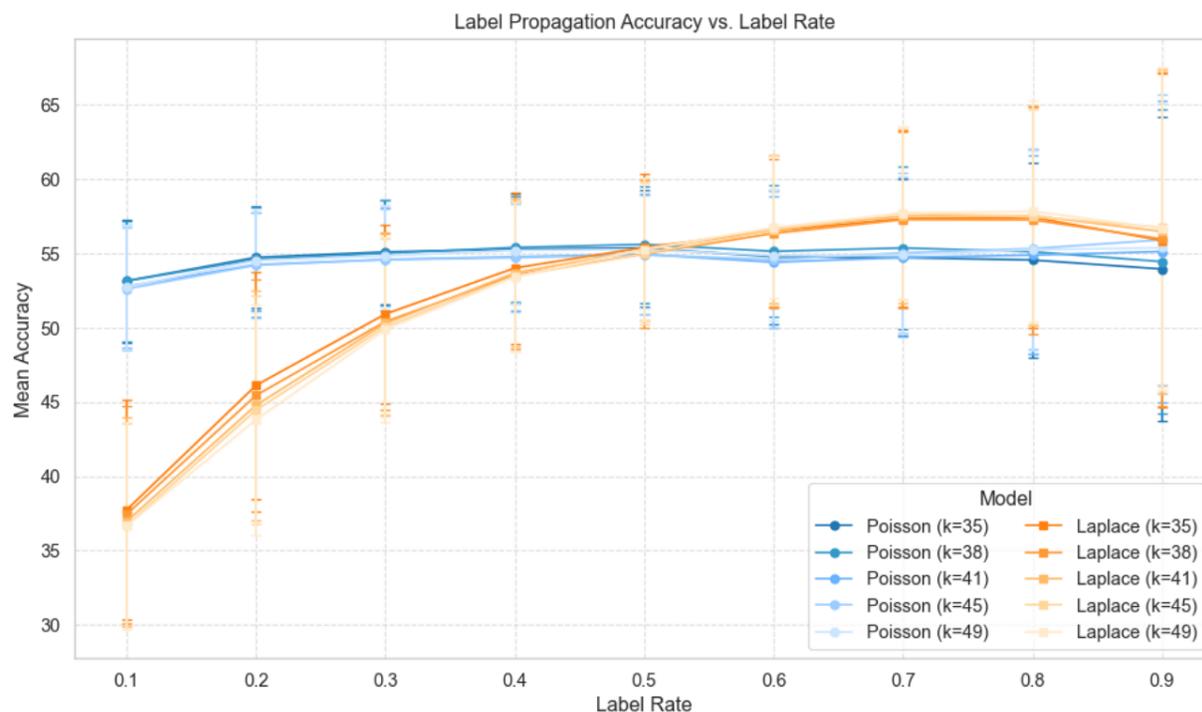

*Figure 8: Study of the evolution of mean accuracy scores for varying label rates r during label propagation experiments for 3 experience levels (Novice, Adept, Advanced). The experiments are computed for Poisson (in blues) and Laplace (in oranges) SSL models, with a number k of considered neighbours in the range [35-49]. The means have been calculated for i = 2000 iterations and the input threshold pair is (0.70, 0.002) as per Figure 7 (left).*

Figure 9 Figure 9 presents the final cluster comparison between the two presented and developed methods of this study. The same data is displayed on bar plots with absolute counts, and on radar plots with proportions relative to the respective clusters. The areas revealed by the radar plots help better visualising overlap in experience levels across clusters.

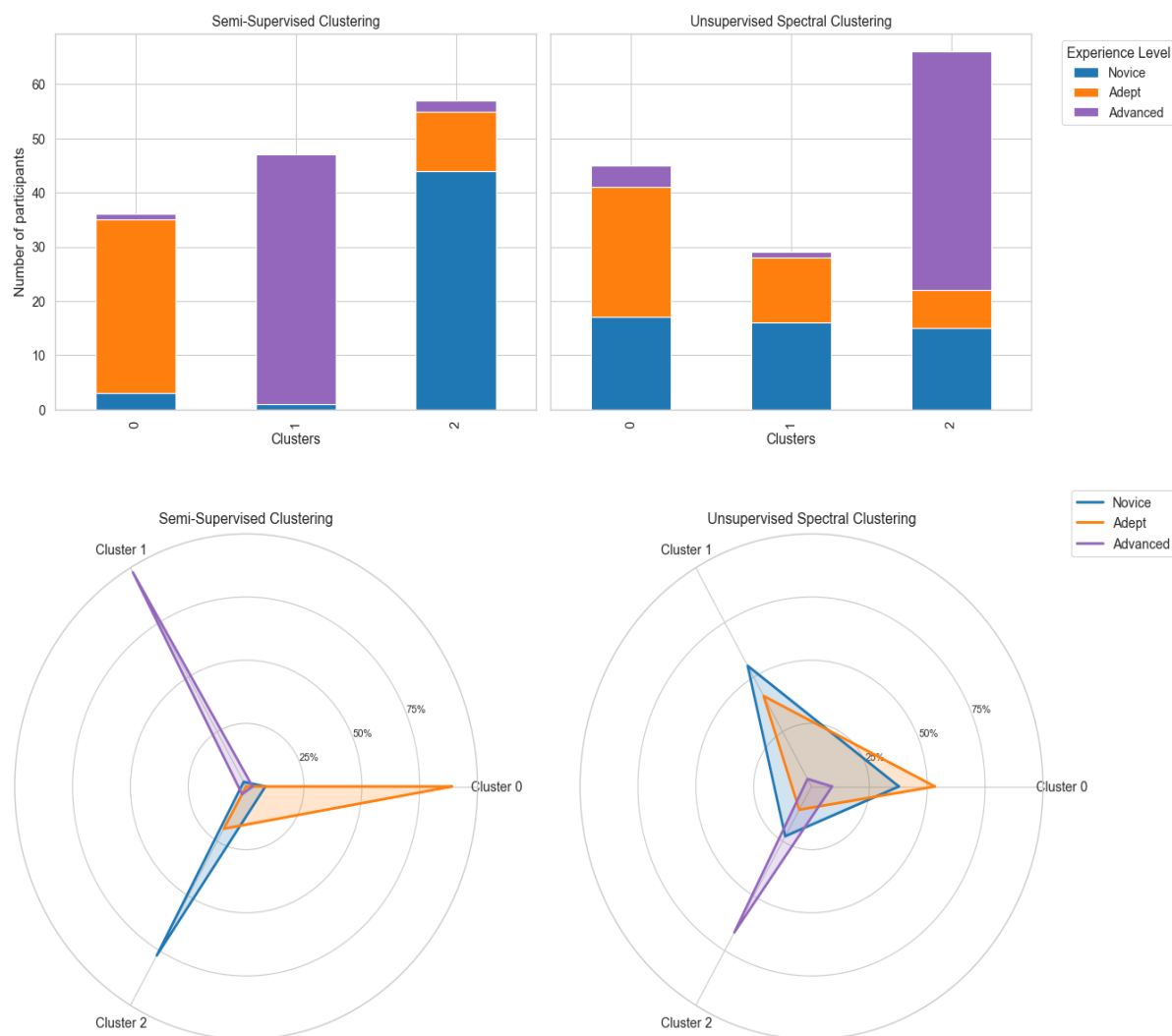

*Figure 9: Cluster comparison between the semi-supervised method and the unsupervised spectral method applied to 3 experience levels (Novice, Adept, Advanced). The bar plots on the top row display each cluster with their absolute counts of participants, and colour-coded for their experience level. The bottom row shows a radar plot of the same data, but displayed in proportion of their respective cluster, with the percentage levels increasing with the radius of each radar.*

We summarise the top 5 high-variance features for each SSL cluster of Figure 9 in Table 1. A complete list of features ranked by decreasing variance for each cluster is available in Supplementary File 1, available to the reader on request.

| Cluster | Feature | Variance |
|---|---|---|
| **0** (Adept) | Evaporation (betweenness) | 0.146 |
| | Average shortest path length | 0.142 |
| | Moist Air (betweenness) | 0.135 |
| | Advection (betweenness) | 0.113 |
| | Weight Dominates Updraft (degree) | 0.086 |
| **1** (Advanced) | Moist Air (right eigenvector) | 0.097 |
| | Dissipation (outdegree) | 0.097 |
| | Bergeron Findeisen (degree) | 0.090 |

| | Precipitation (left eigenvector) | 0.084 |
| --- | --- | --- |
| | Growth Splintering (degree) | 0.081 |
| 2 (Novice) | Evaporation (betweenness) | 0.135 |
| | Rain (degree) | 0.119 |
| | Cloud Formation (left eigenvector) | 0.078 |
| | Precipitation (left eigenvector) | 0.077 |
| | Rain (left eigenvector) | 0.070 |

Table 1: Summary of the top-5 high-variance features for each semi-supervised cluster of Figure 9. The term 'centrality' has been omitted (after betweenness, left/right eigenvector) for better readability.

## Discussion

### Analysis of variance

As displayed in Figure 3, 'experience' has the largest mean $F$-value across all three categorical variables, which means that on average, graph metrics vary significantly between experience levels, compared to variation within each group. This finding aligns with the well-established notion that time, and therefore disciplinary experience, is the most important factor in developing expertise, as demonstrated through various educational research paradigms, such as ACT-R learning (Anderson & Lebiere, 2014; Anderson & Schunn, 2013). The results are accompanied by a marginal p-value of 0.001, indicating total statistical confidence and significance of the mean $F$-value. To answer RQ1 *Among learners' self-assessed experience, academic level, and disciplinary background, which of these characteristics best explains variance in their graphs?*, we conclude that experience is the characteristic best explaining variance in the graphs of the participants. Our subsequent clustering analyses therefore focus on the experience levels to separate the participants across the dataset.

### Unsupervised spectral clustering

The composition of the resulting cluster is shown in Figure 7 (left). Most of the *Advanced* learners (90%, not shown in figure) are grouped in cluster 2, supporting the idea that their differences in conceptual understanding distinguishes them significantly from learners with less experience, namely those in clusters 0 and 1. Interestingly, a few *Novice* and *Adept* learners (respectively 31% and 16%, idem) are also assigned to cluster 2. These individuals may possess strong scientific knowledge or broader interdisciplinary understanding, or they may have underestimated their own disciplinary exposure, leading to them being categorised as less experienced than the *Advanced* group during the categorisation of their self-reported experience. Conversely, the *Advanced* learners in cluster 0 may reflect individuals with low retrieval strength (Bjork & Bjork, 1992), where knowledge is stored but not readily accessible during tasks requiring active recall, such as concept mapping, or learners overstating their academic exposure to the discipline.

One question that remains is whether grouping the *Proficient* and *Expert* learners into an *Advanced* group as argued for and performed by Weihs, Gjerde, et al. (2025) is statistically valid. To answer this, we refer to Figure 7 (right), which results from an investigation of spectral clustering into three clusters, but with the four initial experience levels. A high concentration of *Proficient* and *Expert* learners in the cluster 2 is noticeable, indicating that their merging into an *Advanced* group is favoured by the clustering algorithm. This indicates an overall grouping of learner experience levels into three categories, which aligns with approaches used in physics education (Snyder, 2000; Tong et al., 2023) and other fields of educational research (Bergee, 2005; Wiggins et al., 2002).

A comparison of the left and right figures reveals that both clusters 0 are composed of the same learners, and that the main difference lies between the distribution of *Advanced* learners between clusters 1 and 2: some *Proficient* learners feature in cluster 1 when four experience levels are considered (right), and they could therefore be regarded as 'lower *Advanced*' learners in the three-level nomenclature (left). Both clusters 0 therefore also likely represent the lowest level of experience, almost exclusively containing *Novice* and *Adept* learners in near-equal parts. This in turn suggests that while there is a clear difference when a learner is characterised as *Advanced*, according to Figure 7 (left), the difference between *Novice* and *Adept* is not statistically noticeable. Based on the categorisation of the self-assessed experience of the participants by Weihs, Gjerde, et al. (2025), our study suggests that experience only increases significantly beyond introductory academic exposure to the discipline, with the learners dedicating more time to the field, either through specialisation courses or disciplinarily focused study programmes. According to unsupervised spectral clustering, separating learners into *Novice* and *Adept* is not a sufficient criterion as learners from these two categories feature in all clusters of the three-level nomenclature (Figure 7, left). We therefore suggest that additional or complementary assessment criteria are necessary to better distinguish *Novice* from *Adept* learners in concept mapping tasks. Grouping *Proficient* and *Expert* learners into a group however is confirmed by clustering analysis (Figure 7, right), and beyond a certain level of academic experience, a learner is statistically expected to display higher-level conceptual understanding in their explanation of a discipline.

We can answer RQ2 *Which learner graphs exhibit similarities in terms of structural graph- and node-level properties?* by stating that spectral clustering uncovers a new and complementary partitioning to self-assessed disciplinary experience: the *Advanced* grouping is algorithmically justified, while separating *Novice* and *Adept* learners might not reflect their differences and similarities in conceptual understanding. The clusters in Figure 7 (left) and Figure 7 (right) are amongst the best ways to separate the learners into three groups, based on their similarities in terms of structural graph- and node-level properties of their graphs, considering three and four experience levels respectively.

### Semi-supervised clustering

The investigation of the influence of partial knowledge on the quality of clusters leads to experiments of label propagation. The resulting Figure 8 displays how the accuracy score evolves with different choices of parameters and models. We pick a combination of parameters maximising the accuracy, resulting in a rate of $r_{max} = 0.7$, representing partial knowledge of 70% of the participants' experience in each level, and $k_{max} = 38$ for the choice of the Laplace model. The robustness of these parameters is supported by the accuracy values for both nearby $k$ and $r$. The resulting accuracy is close to 58%, which is comparable with the maximal accuracy of 60% for the unsupervised method. We also note that despite the orange and blue curves diverging after intersecting, the error bars are too large to make either Laplace or Poisson methods significantly different from one another. For the 4-cluster analysis, a similar choice of label rate maximises the mean accuracy at around 45%, which is the highest associated value for the spectral method.

With these parameters, we generate SSL clusters, which we compare to the unsupervised ones in Figure 9. A striking difference lies in how well the SSL method separated the experience levels into respective clusters, with cluster 1-SSL being the *Advanced* cluster, and therefore comparable to cluster 2-spectral. While the clusters 0-spectral and 1-spectral are hard to differentiate, cluster 0-SSL is the apparent *Adept* cluster, and cluster 2-SSL is the *Novice* cluster. This better partitioning of the data draws more attention to the outliers of each cluster: the *Novice* outliers in the *Adept* and *Advanced* SSL-clusters have displayed significantly higher conceptual understanding than expected. Inversely, the *Adept* and *Advanced* learners of cluster 2-SSL show significant shortcoming in their understanding of the discipline. With this approach, our method can also be used to detect outliers overestimating their own experience in a field, and thus is a tool of formative assessment in both classroom and groups comprising learners of different disciplinary experience levels.

From the radar representation of the cluster proportions, the similarities between cluster 1-spectral and cluster 2-spectral become once again apparent. In contrast, the SSL method managed to differentiate what distinguishes *Novice* from *Adept* learners. Reflecting on RQ2, the SSL method allows to surpass the capacities of the unsupervised method to find graph- and node-level criteria to differentiate learners' experience levels. This is also verified for a 4-cluster analysis (not shown in figure), successfully separating all experience groups in distinct clusters. Such result greatly supports the categories used by Weihs, Gjerde, et al. (2025) in their study, and provides statistical evidence for their routine. More generally, the SSL results confirm that categorising learners in three levels of experience has significant relevance in a context of educational research.

To answer RQ3 *To what extent can partial information about learners' experience improve classification into experience levels?*, we conclude that partial information about learner's experience has the potential to greatly improve classification into experience levels. As expected, and as a sign of continuity, running several tests with varying *r* values, the shape of the SSL clusters converges towards that of the unsupervised ones with decreasing labelling rates *r*. Our findings complement other studies in educational research showing that SSL can greatly enhance predictive accuracy in various situations, also in comparison to fully-supervised methods (Kostopoulos et al., 2015; Livieris et al., 2019; Zhang & Hew, 2025). Refining the clusters from our data using SSL routines enables a deeper inquiry about the outlier-learners, but also about the nature of the most significant features characterising each cluster, as investigated by RQ4 below.

Table 1 reveals the metrics most explaining the inter-cluster variance for each SSL cluster, which addresses RQ4 *Which graph features explain the most variance between experience-based clusters?*. The *Novice* group characteristically considers the concept of *Evaporation* as a necessary narrative milestone in their descriptions, and connects a lot of concepts to that of *Rain*. The roles of *Cloud Formation*, *Precipitation*, and *Rain* is also strongly valued as key finishing elements in the description of the life-cycle of a cloud for this group of learners. With increasing experience, the *Adept* group places similar importance on *Evaporation* being an essential crossing concept for any aspect of their explanations; but also on the concepts of *Moist Air*, and *Advection*. In addition, they connect a lot of concepts to the fact that at some point in a droplet's growth, its weight will surpass the updraft forces (leading to the droplet/crystal falling out of its cloud). Lastly, the *Advanced* group places high importance on *Moist Air* as a starting concept, and *Precipitation* as a finishing concept in their descriptions. They also connect a lot of concepts to the *Bergeron Findeisen* process and the concept of *Growth by Splintering*. Finally, they also use *Dissipation* as a starting point to talk about several other concepts. These refined findings with educational potential and impact confirm scholarly output from previous research on conceptual understanding in atmospheric science (Gopal et al., 2004; Rappaport, 2009; Weihs, Euler, et al., 2025). Importantly, we also here show that the most significant graph-derived metrics to distinguish learner groups are node-level ones, as used amongst others by Giabbanelli et al. (2023); Koponen and Pehkonen (2010); Thurn et al. (2020), and recommend all educational research using network analysis on conceptual data to include centrality measures in their analysis.

Our results shed light on the conceptual understanding of different experience groups in a discipline, and enable a further investigation of the features that characterise increasing levels of experience. Such insights could be used in teaching and learning context, to assess the main knowledge structures of a group of learners, to reveal misconceptions, and to structure a learning progression towards a more advanced and elaborate understanding of the discipline and its key concepts. We have shown the performance of both unsupervised and SSL clustering models under parametrisation, and demonstrated that SSL methods could improve and refine the grouping of participants in educational research studies. We lastly suggest that the application of our methodology could benefit the educational research informing teaching and learning in any STEM discipline, and hope to inspire the readership to use such techniques in their fields of activity.

## Acknowledgements

The research is supported by the Norwegian Centre for Integrated Earth Science Education iEarth (Norwegian Agency for International Cooperation and Quality Enhancement in Higher Education grant #101060).

## Declaration of interest statement

The authors report there are no competing interests to declare.

# Appendix

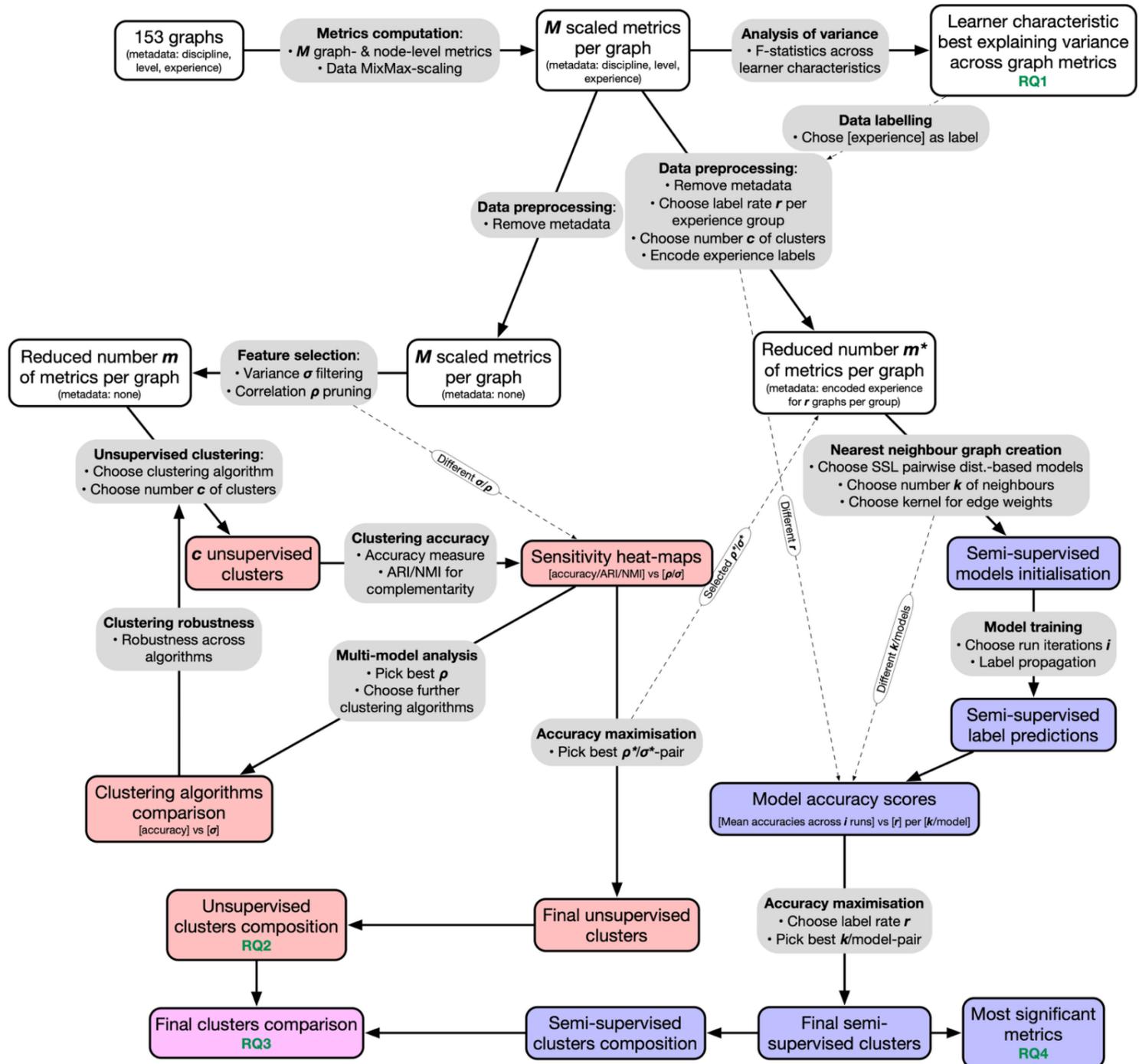

*Figure 10: Full work pipeline describing the unsupervised (in red) and semi-supervised (in blue) clustering analyses of the participants data according to various graph- and node-level metrics. The outputs addressing the four research questions of this study are highlighted with the mentions 'RQx' (in green), and the procedure comparing both clustering methods features in pink. The symbols used ($c$, $r$, $i$, $k$, $\rho$, $\sigma$ (for the parameter $\sigma^2$)) are addressed in the Methods section. The dotted lines and associated labels address the changing inputs of variables or models in different processes of the workflow.*